\documentclass[aps,prl,amsmath,amssymb,prb,superscriptaddress,twocolumn,showpacs,longbibliography,floatfix,notitlepage]{revtex4-1}
\usepackage{graphicx}
\usepackage{dcolumn}
\usepackage[colorlinks=true,linkcolor=blue ,citecolor=blue,urlcolor=blue]{hyperref}
\usepackage{multirow}
\usepackage[usenames,dvipsnames]{xcolor}
\usepackage{soul}
\usepackage{braket}
\usepackage{bm}
\usepackage{nicefrac}
\usepackage{siunitx}
\usepackage{color}
\usepackage[utf8]{inputenc}   
\usepackage{upgreek}

\usepackage{float}

\newcommand{\VEC}[1]{\boldsymbol{#1}}

\begin{document}

\title{Transverse Transport in Two-Dimensional Relativistic Systems with Non-Trivial Spin Textures}

\author{Juba Bouaziz}
\email{j.bouaziz@fz-juelich.de}
\affiliation{Peter Gr\"{u}nberg Institut and Institute for Advanced Simulation, Forschungszentrum J\"{u}lich \& JARA, D-52425 J\"{u}lich, Germany}
\author{Hiroshi Ishida}
\affiliation{College of Humanities and Sciences, Nihon University, Sakura-josui, Tokyo 156-8550, Japan}
\author{Samir Lounis}
\affiliation{Peter Gr\"{u}nberg Institut and Institute for Advanced Simulation, Forschungszentrum J\"{u}lich \& JARA, D-52425 J\"{u}lich, Germany}
\affiliation{Faculty of Physics, University of Duisburg-Essen, 47053 Duisburg, Germany}
\author{Stefan Bl\"ugel}
\affiliation{Peter Gr\"{u}nberg Institut and Institute for Advanced Simulation, Forschungszentrum J\"{u}lich \& JARA, D-52425 J\"{u}lich, Germany}

\date{\today}

\begin{abstract}
Using multiple scattering theory, we show that the generally accepted expression of transverse 
resistivity in magnetic systems that host skyrmions, given by the linear superposition of the 
ordinary (OHE), the anomalous (AHE) and the topological Hall effect (THE), is incomplete and 
must be amended by an additional term, the "non-collinear" Hall effect (NHE).
Its angular form is determined by the magnetic texture, the spin-orbit field of the electrons, 
and the underlying crystal structure, allowing to disentangle the NHE from the various other 
Hall contributions.  Its magnitude is proportional to the spin-orbit interaction strength. 
The NHE is an essential term required for decoding two- and three-dimensional spin textures 
from transport experiments. 
\end{abstract}

\maketitle

The electronic transport is paramount in condensed matter physics. It permits the probing of 
electronic and magnetic properties of solids by electrical means and exhibits a plethora 
of exciting phenomena. One of them is the Hall effect, the response measured transversely 
to an electric current and a perpendicularly applied magnetic field. It manifests itself 
in systems with broken time-reversal symmetry, either due to an external magnetic 
field~\cite{Hall:1879} leading to the ordinary Hall effect (OHE) or due to the presence 
of spontaneous magnetization, the latter is known as the anomalous Hall effect (AHE)~\cite{Luttinger:1958,nagaosa:2006,Nagaosa:2010}. 
The topological Hall effect (THE) is an additional contribution {attributed to a Berry 
phase~\cite{Berry:1984} acquired by the electrons following adiabatically a smoothly 
varying non-coplanar magnetization texture~\cite{Taguchi:2001,Bruno:2004,Tatara:2007}.} 
The linear superposition of the three  contributions
yields the commonly accepted expression for the Hall resistivity~\cite{Neubauer:2009,lee:2009,kanazawa:2011,li:2013,hamamoto:2015,yin:2015,lado:2015,gobel:2017,gobel:2018,nakazawa2018topological}:
\begin{equation}
\rho^\text{H}_\text{Tot}= \rho^\text{OHE} + \rho^\text{AHE} + \rho^\text{THE}\quad,
\label{resist_supr_lin}    
\end{equation}
where the first term scales linearly with the applied external magnetic field, the second $(\rho^\text{AHE})$ 
is linear in terms of the magnetization, and the last term $(\rho^\text{THE})$ is proportional 
to the emergent magnetic field~\cite{Taguchi:2001,Bruno:2004,Tatara:2007}, that is in turn 
proportional to the topological skyrmion number  defined by the sum of all solid angles 
$\VEC{m}_i\cdot(\VEC{m}_j\times\VEC{m}_k)$ of the magnetic moments at three different 
sites $i,j,k$. Equation~\eqref{resist_supr_lin} was first introduced in the seminal work 
of Neubauer \textit{et al.}~\cite{Neubauer:2009}, in which the connection between 
$\rho^\text{THE}$ and emergence of the skyrmion lattice ($A$ phase) in B20 MnSi was 
established. Since Eq.~\eqref{resist_supr_lin} allows a fully electric detection and quantification~\cite{Neubauer:2009,schulz2012emergent,meng2019observation,soumyanarayanan2017tunable,spencer2018helical,raju2019evolution,liang2015current,schulz2012emergent,kanazawa2011large,jiang2020concurrence,ludbrook2017nucleation} of magnetic 
skyrmions, \eqref{resist_supr_lin} is of crucial importance for the characterization 
of skyrmions, especially compact skyrmions smaller than the resolution limit of 
imaging techniques. 

In recent years, however, especially for magnetic films and multilayers with interfaces 
of heavy materials having large spin-orbit interactions, doubts have been raised about 
the simple relationship between the Hall resistivity and the skyrmion count~\cite{Wysocki:2020}.
For instance, a quantitative analysis on Ir/Fe/Co/Pt multilayers 
shows much larger experimental topological-Hall resistivities than expected 
{from the skyrmion density measured by magnetic force microscopy}, which was partly 
explained by {contributions of} worm-like spin textures with a non-vanishing 
topological charge~\cite{raju2019evolution}. In addition, in the  weak coupling limit 
{of the moving electron spin to the underlying magnetic texture} small vortex-like 
textures without topological charge are able to produce a finite topological Hall effect, 
which vanishes for larger spin textures~\cite{denisov2017nontrivial}, indicating 
its non-topological nature.

In this letter we evaluate the transverse resistivity of a two-dimensional {(2D)} 
ensemble of non-collinear magnetic  atoms interacting through a relativistic electron 
gas by means of  multiple scattering theory. Treating spin-orbit interaction (SOI) 
and non-collinear magnetism (NCM) on the same footing we demonstrate that the Hall 
resistivity tensor  \eqref{resist_supr_lin} needs to be amended by an additional term,
the {\em non-collinear} Hall effect (NHE):
\begin{eqnarray}
\label{NHE_contribution}
\rho^\text{NHE} & = & \sum_{ijl} \rho^\text{NHE}_{ijl} 
\left[ (\right.\VEC{\mathcal{D}}_{ij}\cdot\VEC{m}_i)(\VEC{m}_j\cdot\VEC{m}_{l}) \\ 
\nonumber
& + & (\VEC{\mathcal{D}}_{ij}\cdot\VEC{m}_{j})(\VEC{m}_{i}\cdot\VEC{m}_{l})
- (\VEC{\mathcal{D}}_{ij}\cdot\VEC{m}_{l})(\VEC{m}_i\cdot\VEC{m}_j)\left.\hspace{-1mm}\right],
\end{eqnarray}
looming from the interference of both {the SOI and the NCM}. $\VEC{m}_i$ indicates the unit vector of the magnetic moment at site $i$. $\rho^\text{NHE}_{ijk}$ depends  like the  Dzyaloshinskii-Moriya interaction (DMI)~\cite{Dzyaloshinsky:1958,Moriya:1960} on the spin-flip components of the relativistic Green function of 
the electron gas~\cite{bouaziz2017chiral}, and thus correlates to the SOI, depends on the electronic structure. The orientation of the unit vector $\VEC{\mathcal{D}}_{ij}=-\VEC{\mathcal{D}}_{ji}$ depends on the direction of the vector connecting the sites $i$, $j$, the crystal symmetry coupling  orbital with spin degrees and follows the symmetry rules of the DMI.

Having the description of surfaces, films and multilayers in mind we model the 
electronic structure by a 2D  electrons gas under Rashba-, Dresselhaus- or 
Weyl-type SOIs. The magnetic atoms are represented by a collection of localized 
spins with an $sd$-like coupling to the free-electrons~\cite{Fransson:2015}. 
The scattering off the magnetic sites is taken in the weak-coupling limit~\cite{Denisov:2016,denisov2017nontrivial}, 
whereby not only adiabatic but also non-adiabatic effects on the resistivity are considered. 

We show that the NHE can assume significant contributions in non-collinear magnets, 
that it occurs in two- and three-dimensional spin textures and that it can be used 
to discriminate Bloch- from N\'eel-type skyrmions. We suggest an experimental 
protocol to disentangle the NHE from other contributions at play. Considering 
the micromagnetic limit of \eqref{NHE_contribution} we recover the recently 
conjectured chiral Hall effect~\cite{Lux:2020}, and unravel its  microscopic 
origin. 

\begin{figure*}
    \centering
	\includegraphics[width=1.0\textwidth]{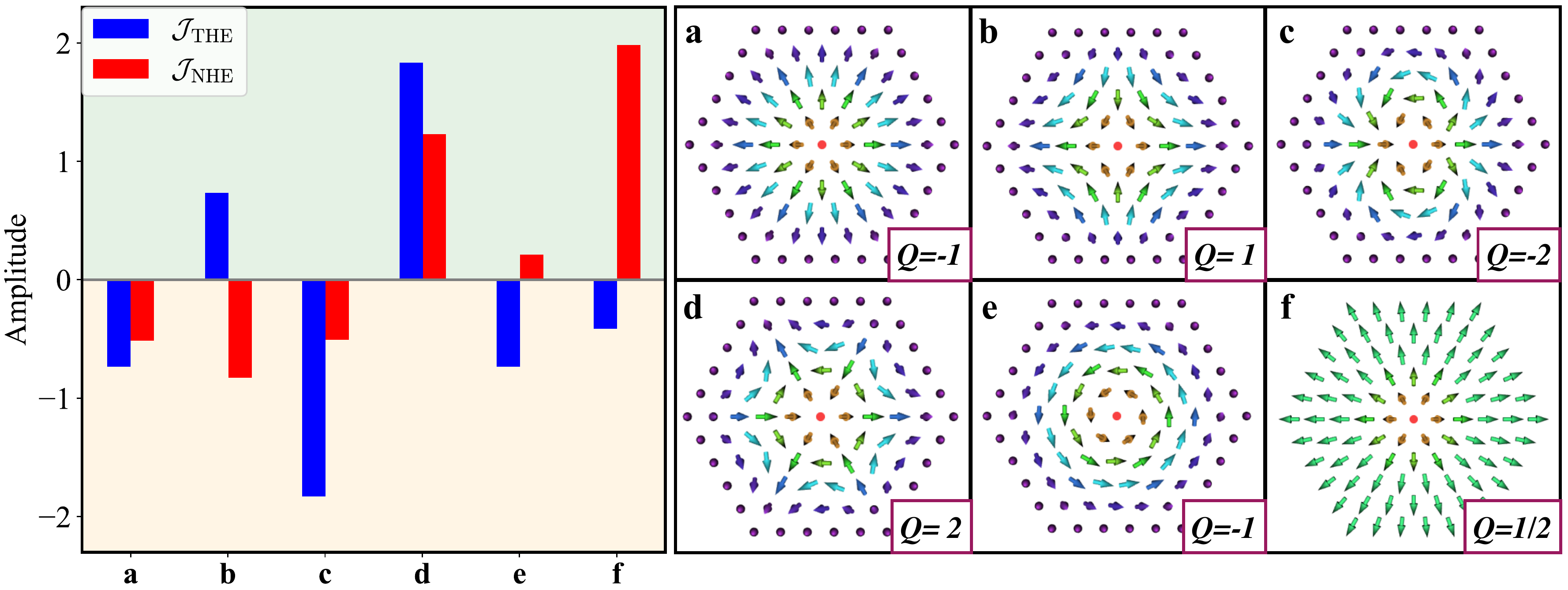}
	\caption{The chart in the left panel describes the size of the topological 
	(blue) and non-collinear (red) Hall current for six topological spin textures 
	calculated for Rashba-type SOC with spin-orbit strength of $\mathcal{R} = 25\%$. 
	The latter are shown in the right panel: a) {N\'eel-type} skyrmion, b) anti-skyrmion, 
	c) bi-skyrmion, d) anti-bi-skyrmion, e) Bloch-type skyrmion, f) meron. 
	We consider clusters containing $91$ magnetic sites with a nearest neighbor 
	distance of $d=3$ \AA. smearing $\Gamma = 0.2$~eV 
	is used {in the vicinity of the Fermi energy} to account for eventual temperature or 
	disorder effects.} 
	\label{THE_NHE_skyrmion_types}
\end{figure*}

The starting point of the derivation of \eqref{NHE_contribution} is the multiple 
scattering theory applied to a collection of sites forming a complex magnetic 
configuration as of a nano-skyrmion shown in Fig.~\ref{THE_NHE_skyrmion_types}a. 
An incident free-electron in the state $\ket{\VEC{k}\sigma}$ with an energy $\mathcal{E}_{\VEC{k}} = \hbar k^2/2m$ 
($k$ and $m$ being the wave number and electron mass, respectively) carrying a 
spin $\sigma$ scatters {elastically at a} collection of localized potentials~\cite{Fiete:2003,Juba:2016} 
{into a state $\ket{\VEC{k}^\prime\sigma^\prime}$ with ${k}={k}^\prime$, and 
$\VEC{k}$ and $\VEC{k}^\prime$ including a scattering angle $\phi=\angle\VEC{k} \VEC{k}^\prime$}.
In the far-field limit, the resulting wave function consists of a linear superposition 
of the incoming wave and the outgoing scattered one given by~\cite{Juba:2016,Denisov:2016,denisov2017nontrivial}:
\begin{equation}
\psi_{\VEC{k}\sigma}(\vec{r}) = e^{i\VEC{k}\VEC{r}}\ket{\sigma} +\sum_{\sigma^\prime} 
\frac{e^{ikr}}{\sqrt{r}}f_{k,\sigma^\prime\sigma}(\phi)\ket{\sigma^\prime}\quad.
\end{equation}
$\ket{\sigma}$ is the eigenstate of $\hat{S}_{z}$. $f_{k,\sigma^\prime\sigma}(\phi)$ 
represents the scattering amplitude, a central quantity within our {scattering} approach, 
which is defined via the differential cross section
\begin{equation}
\frac{d\Sigma^\text{H}}{d\phi} = \sum_{\sigma\sigma^\prime} |f_{k,\sigma^\prime\sigma}(\phi)|^2\quad. 
\label{assym_scatt_hall}
\end{equation}
directly related to the Hall current $\mathcal{J}_\text{H} = k\Sigma^\text{H}$ (see Supplementary Material~\cite{supp_cite}). 
Following  standard scattering theory, $f_{k,\sigma^\prime\sigma}(\phi)$ 
is evaluated by the Lippmann-Schwinger equation~\cite{Juba:2016,Denisov:2016} employing the {relativistic} 
Green function of the propagating electron, and including the multiple scattering events experienced 
by the incident electron within the skyrmion via the scattering matrix 
$\underline{{\mathcal{T}}}$. {In the weak coupling regime~\cite{Denisov:2016,denisov2017nontrivial}, 
the ($\underline{{\mathcal{T}}}$)-matrix is computed using the second Born approximation}.
Inside the skyrmion, the electron propagates under a spin-orbit field in a structure asymmetric 
environment~\cite{bychkov1984oscillatory,bouaziz2018impurity}. This approach allows to establish 
a transparent relation between the scattering cross section and the orientation of the magnetic 
moments in the presence of relativistic effects (as shown in Supplementary Material~\cite{supp_cite}). 
{The presence of the combined effect of SOI and NCM results in a finite asymmetric 
(right/left) scattering in the transverse direction, \textit{i.e.} a finite Hall effect.} 
The Hall current $\mathcal{J}_\text{H}$ is then tied back to the resistivity employing 
the Boltzmann equation (see Supplementary Material~\cite{supp_cite}). Finally, the electric current is considered 
with respect to the limit of small external electric fields. Since in this case the electron 
transport is dominated by the Fermi surface contributions~\cite{Sinova:2015}, the current is 
calculated at the Fermi energy. 

\begin{figure*}
    \centering
	\includegraphics[width=1.0\textwidth]{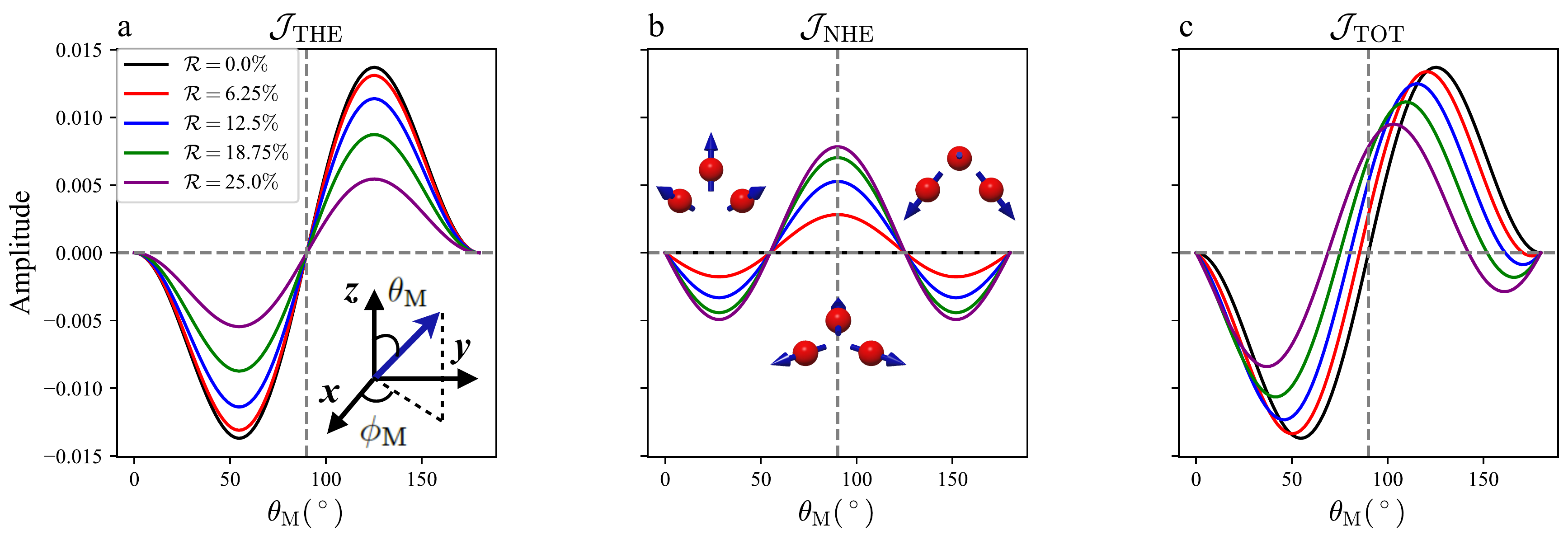}
	\caption{
	Evolution of the scattered current for a magnetic trimer in an equilateral
	geometry as a function of the opening angle $\theta_\text{M}$ of the magnetic 
	moments and of the spin-orbit strength $\mathcal{R}$. a) The THE contribution 
	to the current determined by the scalar spin chirality. b) NHE current originating 
	from relativistic effects, its  non-trivial angular dependence is given in 
	Eq.~\eqref{NHE_contribution}. c) Total Hall current emerging from three spin processes, 
	which present an non-symmetric dependence for $\theta_\text{M}\rightarrow\pi-\theta_\text{M}$.}
	\label{Panel_THE_NHE_TOT}
\end{figure*}
In order to illustrate and identify the contribution of the NHE to the Hall signal, 
we consider a magnetic trimer, which is the smallest nano-structure 
capable of generating a finite scalar spin chirality~\cite{dos2016chirality,bouaziz2018impurity}. 
The three magnetic atoms form an equilateral triangle separated by a distance of ${d} = 3$~\AA. 
The Fermi energy is set to $\varepsilon_\text{F} = 0.8$~eV, 
which coincides with the Fermi energy at metallic surfaces~\cite{reinert2001direct}. 
We consider the weak scattering limit by setting the spin-dependent scattering phase 
shift to $\delta^\uparrow = \frac{7\pi}{8}$ and $\delta^\downarrow = \frac{5\pi}{8}$ for 
spin-up and -down electrons, respectively. The evolution of the outgoing Hall current as 
function of spin non-collinearity expressed in terms of the opening angle of the magnetic 
moments $\theta_\text{M}$, and the spin-orbit field of the electrons described here by 
the Rashba-model with a spin-orbit strength expressed by the ratio $\mathcal{R} = {{k}_\text{so}}/{k_\text{F}}$ 
of the spin-orbit wave vector ${k}_\text{so}$ and the Fermi wave vector $k_\text{F}$ 
is depicted in Fig.~\ref{Panel_THE_NHE_TOT}. The associated azimuthal angle $\phi_\text{M}$ 
of the moments at each corner of the triangle changes from corner to corner by $120^\circ$, 
\textit{i.e.}\ $\phi_\text{M}=\{\pi/2,7\pi/6,11\pi/6\}$. Three different quantities are 
shown: the THE, the NHE and their sum. The value of $\mathcal{R}$ can be engineered to 
reach rather large magnitudes by tuning the Rashba spin splitting 
{$\propto {k}_\text{so}$}~\cite{ishizaka2011giant,Ast:2007,frantzeskakis2010tuning}.

The THE contribution shown in Fig.~\ref{Panel_THE_NHE_TOT}a follows the scalar spin 
chirality $\chi_{ijk} = \VEC{m}_i\cdot(\VEC{m}_j\times\VEC{m}_k)$, it  maximises at 
$\theta_\text{M} = 60^\circ$, $120^\circ$ and vanishes at the ferromagnetic 
($\theta_\text{M} = 0^\circ$, $180^\circ$) and coplanar ($\theta_\text{M} = 90^\circ$) 
spin configuration. Interestingly, the THE current is altered by the SOI as it decreases 
for larger values of $\mathcal{R}$, a closer inspection of $\mathcal{J}_\text{THE}$ 
shows that it depends only on the spin-conserving components of the Rashba Green function. 
In the limit $\mathcal{R} \ll 1$, the latter decreases quadratically upon 
increasing spin-orbit strength. On the other hand, the NHE shown in Fig.~\ref{Panel_THE_NHE_TOT}b 
reaches its highest value when $\theta_\text{M} = 90^\circ$. This can be attributed 
to a larger scalar product $(\VEC{\mathcal{D}}_{ij}\cdot\VEC{m}_{i})$ since 
$\VEC{\mathcal{D}}_{ij}$ is locked inside the $(xy)$-plane. It also results 
in a symmetric signal when ${m}_{z}\rightarrow -{m}_{z}$. The strength of the 
NHE increases linearly for small values of $\mathcal{R}$, and is determined 
by the spin off-diagonal components of the Green function (Supplementary Material~\cite{supp_cite}), 
similarly to the DMI~\cite{bouaziz2017chiral}. 

In order to experimentally distinguish between the THE and NHE we make use 
of the observation that by including the NHE in the overall picture, the Hall 
signal becomes asymmetric when $\theta_\text{M}\rightarrow\pi-\theta_\text{M}$ 
(see Fig.~\ref{Panel_THE_NHE_TOT}c). Thus, considering a Hall setup for a skyrmion 
system, two sets of measurements with opposite orientations of the major axis 
of the crystal should be performed. The anti-symmetric part of the signal is 
the THE, while the symmetric one leads to the NHE.


A remarkable property of the NHE is its dependence on the crystal structure, which 
underlines its deep relationship with the SOI. The vector $\VEC{\mathcal{D}}_{ij}$ 
provides a direct connection between the electronic structure of the material and 
the NHE (see Supplementary Material~\cite{supp_cite}). Therefore, Eq.~\eqref{NHE_contribution} can 
be expressed for the well-known 2D spin-orbit Hamiltonians. The different expressions
of $\VEC{\mathcal{D}}^{\text{2D}}_{ij}$ for the Rashba~\cite{bychkov1984oscillatory}, 
Dresselhaus~\cite{winkler2003spin} and Weyl Hamiltonians~\cite{Wan:2011} are given 
in Table~\ref{all_ham_Dij}. This offers an additional perspective for the engineering 
of the functional dependence of the NHE on the magnetization direction using the 
symmetry properties of the electronic band structure (crystal structure),
besides tuning its magnitude by changing the coupling parameter as shown in 
Fig.~\ref{Panel_THE_NHE_TOT}. Note also that once the NHE can be analyzed separately 
from the overall Hall response, it could be used as a means to study the band structure 
and spin momentum locking based on all-electric measurements. The relation between 
the NHE and the crystal structure is an illustration of Neumann's principle, which 
dictates that the point  group {symmetry} of a system is reflected in its physical 
quantities~\cite{bradley2009mathematical,vsmejkal2018symmetry,Seemann:2015}.

\begin{table}
\caption{Generic Rashba, Dresselhaus and Weyl Hamiltonians in $\VEC{k}$-space,
		and their corresponding expressions for $\VEC{\mathcal{D}}_{ij}$ defining the 
		NHE resistivity. $\mathcal{\phi}_{ij}$ is the angle between the $\VEC{r}_{ij}$ 
		(bond vector connecting $i$ to $j$) and the $x$-axis (see Supplementary Material~\cite{supp_cite}).} 
        \label{all_ham_Dij}
	\begin{ruledtabular}
		\begin{tabular}{lcc}
			 System &  Hamiltonian  &  $\VEC{\mathcal{D}}_{ij}$    \\
		    \hline		    
			 Rashba      & $\alpha(k_{y}\boldsymbol{\sigma}_{x}-\boldsymbol{\sigma}_{y}k_{x})$  
			 & $(\phantom{-}\sin\mathcal{\phi}_{ij},-\cos\mathcal{\phi}_{ij})$ \\
			 Dresselhaus & $\beta (k_{y}\boldsymbol{\sigma}_{y}-\boldsymbol{\sigma}_{x}k_{x})$  
			 &  $(-\cos\mathcal{\phi}_{ij},\phantom{-}\sin\mathcal{\phi}_{ij})$ \\
			 Weyl        & $\gamma(k_{y}\boldsymbol{\sigma}_{y}+\boldsymbol{\sigma}_{x}k_{x})$  
			 & $(\phantom{-}\cos\mathcal{\phi}_{ij},\phantom{-}\sin\mathcal{\phi}_{ij})$  \\
		\end{tabular}
	\end{ruledtabular}
\end{table} 
After expanding Eq.~\eqref{NHE_contribution} in terms of the components of the 
magnetization, and plugging in the various expressions of $\VEC{\mathcal{D}}^{\text{2D}}_{ij}$ 
provided in Table~\ref{all_ham_Dij}, one yields an expression of the non-collinear 
Hall resistance: 
\begin{equation}
\begin{split}
\rho^\text{NHE}_\text{2D} & = \sum_{ijl}\rho^{\text{2D}}_{ijl}\, \VEC{\mathcal{F}}_{ijl}(\VEC{m})
\left(\underline{\mathcal{S}}^{\text{2D}}\cdot\hat{\boldsymbol{r}}_{ij}\right)\quad. 
\end{split}
\end{equation}
$\VEC{\mathcal{F}}_{ijk}(\VEC{m})$ contains {the angular forms of the magnetization}
(for details see Supplementary Material~\cite{supp_cite}). $\hat{\boldsymbol{r}}_{ij}=(\VEC{r}_{i}-\VEC{r}_{j})/
|\VEC{r}_{i}-\VEC{r}_{j}|$, $\VEC{r}_{i}$ defining the position of the magnetic site $i$. 
The matrix $\underline{\mathcal{S}}^{\text{2D}}$ reflects the symmetry of the underlying 
spin-orbit model as:
\begin{equation}
\underline{\mathcal{S}}^\text{R} = i \boldsymbol{\sigma}_{y}\, ,\quad 
\underline{\mathcal{S}}^\text{D} = - \boldsymbol{\sigma}_{z}\, ,\quad
\underline{\mathcal{S}}^\text{W} =  
\boldsymbol{\sigma}_{0}\quad ,
\end{equation}
where $\boldsymbol{\sigma}$ are the conventional Pauli matrices, and $\boldsymbol{\sigma}_{0}$ 
is the unity matrix. R, D and W stand for Rashba, Dresselhaus and Weyl, respectively. 
For the Rashba and Dresselhaus Hamiltonians describing polar systems the form of 
$\underline{\mathcal{S}}^\text{2D}$ is anti-symmetric, while it has a symmetric 
form for the Weyl Hamiltonian. A similar behavior was also identified for the DMI in 
Refs.~\cite{kikuchi2016dzyaloshinskii,kim2013chirality,Freimuth:2017}. 

We turn now to the discussion of the signature of complex magnetic textures~\cite{gobel2020beyond} 
on the NHE. Considered are: Néel- and Bloch-type skyrmions~\cite{Nagaosa:2013}, anti-skyrmions, 
merons~\cite{lin2015skyrmion}, and higher-order skyrmions~\cite{gobel2020beyond}. These topological 
entities are described by the topological charge ($Q=-m$), the vorticity ($m$) and helicity ($\gamma$) 
relating the azimuthal angle of the magnetization, ${\Phi} = m\,\phi_\text{M}+\gamma$, to azimuthal 
angle $\phi_\text{M}$ of the lattice~\cite{Nagaosa:2013}. We consider skyrmions with a small 
radius of $R_\text{sk} = 1.5$~nm, as typical for skyrmions created in magnetic transition-metal 
monolayers at heavy metal interfaces, \textit{e.g.} 
Pd/Fe/Ir(111)~\cite{romming2013writing,crum2015perpendicular,fernandes2018universality}.

Fig.~\ref{THE_NHE_skyrmion_types} depicts  the topological and non-collinear Hall currents for 
six different spin textures in systems with Rashba spin-orbit field {(the results for the 
Dresselhaus and Weyl Hamiltonians are shown in Supplementary Fig.~$3$)}. {Since in the adiabatic 
limit of transport the THE contribution is proportional to the topological charge, Néel- (a) and 
Bloch-type (e) skyrmions} exhibit the same value of $\mathcal{J}_\text{THE}$~\cite{denisov2017nontrivial}. 
The same holds true for the other spin textures, the anti-skyrmion (b) has the opposite charge 
and subsequently the opposite sign of $\mathcal{J}_\text{THE}$, the bi-skyrmion has twice the 
charge and doubles  the magnitude of the current.

On the other hand, the NHE  is found to be non-negligible, and even colossal in some cases without 
any apparent connection to the spin chirality or {the} underlying topology of the magnetic texture. 
It is also notable that the NHE acquires a different value for the Néel (a) and Bloch (e) skyrmions, 
enabling to discriminate  the two species. This difference is due to the NHE being a relativistic 
effect breaking the $\mathcal{SU}(2)$ rotation symmetry.

A pattern emerges from Fig.~\ref{THE_NHE_skyrmion_types}, the NHE is large for {textures} $b$ and $d$, 
hence it is increased by a negative vorticity ($m<0$). At the same time, it is notably reduced 
for the Bloch skyrmion (e), suggesting that it diminishes in the presence of a finite helicity 
$\gamma \ne 0$. The latter argues in favor of the NHE being rather weak or insignificant in bulk 
systems with Bloch skyrmions such as MnSi. Finally, as observed in Fig.~\ref{Panel_THE_NHE_TOT}b
the NHE is largest when spins are locked in-plane. Thus, the drastic increase of the NHE comparing 
the meron (f) to the Néel skyrmion (a) can be inferred to a larger number of magnetic sites lying 
within the {$(xy)$-plane containing the electron gas}. 

In the limit of a slowly varying magnetic texture, \textit{i.e.}\,in the adiabatic 
limit, the Hall resistivity can be expressed in terms of the magnetization density 
and its gradients~\cite{Bruno:2004,Schulz:2012}. Taking this limit and expanding 
Eq.~\eqref{NHE_contribution} to second order in the magnetization gradients, we 
provide a general micromagnetic form of the Hall resistivity, which incorporates 
new anisotropic contributions due to the SOI (see Supplementary Material~\cite{supp_cite}): 
\begin{eqnarray}
\label{Micromagn_all}
\nonumber\rho^\text{H} & = &\sum_{\substack{\alpha\ne\beta}}\rho^\text{THE}_{\alpha\beta}\,\VEC{m}\cdot
\left(\partial_{\alpha}\VEC{m}\times\partial_{\beta}\VEC{m}\right)\\
& + & \sum_{\substack{ijl}}\rho^\text{NHE}_{ijl}\left(\VEC{m}\cdot\VEC{\mathcal{D}}_{ij}\right) + \sum_{\alpha\beta}\rho^\text{NHE}_{1,\alpha\beta}\,\partial_{\alpha}{m}^\beta\\ 
\nonumber& + & \sum_{\substack{\alpha\beta\gamma}}\rho^\text{NHE}_{2,\alpha\beta\gamma}
\,\partial_{\alpha}\partial_{\beta}{m}^{\gamma} 
+ \sum_{\substack{\alpha\beta\gamma}}\rho^\text{NHE}_{3,\alpha\beta\gamma}
\,(\partial_{\alpha}\VEC{m})(\partial_{\beta}\VEC{m})m^{\gamma}\quad, 
\end{eqnarray}
{with $\alpha,\beta,\gamma\in\{x,y,z\}$.} The first term on the right-hand side 
consists of the topological charge density known from the THE~\cite{Bruno:2004,Schulz:2012}. 
Four additional terms are derived: The second term is linear in the magnetization, 
nonetheless, it originates from a three-site scattering process. It leads to an
antisymmetric contribution to the Planar Hall effect, which has been measured in 
the Heusler alloy Fe$_3$Si~\cite{Muduli:2005}. The term linear in the gradient 
of the magnetization we identify as the recently predicted chiral Hall 
effect~\cite{Lux:2020}. The fourth contribution involves the curvature of $\VEC{m}$. 
It is expected to be a non-negligible component for strongly inhomogeneous magnetization 
fields or when $\VEC{m}$ is constricted in curved geometries~\cite{streubel2016magnetism,rybakov2019magnetic}. 
Finally, the last element of Eq.~\eqref{Micromagn_all} has a form similar to the 
THE as it involves the magnetization and a product of its gradients. Which 
components $\{\alpha,\beta,\gamma\}$ contribute are purely determined by the 
SOI-dependent form $\VEC{\mathcal{D}}_{ij}$. In analogy, similar higher order 
gradient/anisotropic corrections may also emerge at the level of the orbital 
magnetisation in presence of chiral spin textures~\cite{lux:2018}. 

The result of Eq.~\eqref{Micromagn_all} is another important outcome of this 
letter. It displays in a clear fashion that when relativistic effects are accounted 
for (even in the adiabatic limit), no topological spin texture is necessary to generate 
a non-conventional Hall effect, \textit{i.e.}\ a Hall effect that is neither proportional 
to the magnetization nor to the external field. The derived formula incorporates various 
terms of different orders that could potentially compete or collectively contribute 
to the Hall signal. As a result, this renders the one-to-one correspondence between 
the spin texture and Hall measurement rather complex and intricate. 

To conclude, we derived  contributions to the transverse resistivity 
emerging from three-site scattering processes at magnetic atoms due to electrons 
subject to SOI and structure inversion asymmetry and   showed that the linear 
superposition of the OHE, AHE, and THE (Eq.~\eqref{resist_supr_lin}) must be 
extended by a new contribution, the non-collinear Hall effect (NHE). The new 
Hall effect has far-reaching consequences, \textit{e.g.}: (i) Its magnitude 
and angular form  can be engineered by tuning the electronic band structure. 
(ii) It can give rise to substantial Hall responses in compensated magnets, 
which we conjecture is the reason for the substantial Hall response observed 
in the kagome magnets Mn$_3$Sn and Mn$_3$Ge~\cite{nayak2016large,Busch:2020}, 
where the NHE can also be easily disentangled from the THE since non-coplanar 
spins are located in the atomic plane. (iii) It can resolve the nature of 
topological spin-textures as our comparison to the THE showed. Finally, 
in the micromagnetic limit, the NHE translates into a superposition of 
different terms, including contributions proportional to the curvature 
of the magnetization, which could be of particular interest in systems 
hosting magnetic Hopfions~\cite{rybakov2019magnetic,grytsiuk2020topological}. 
All these findings open a new vista for the analysis of Hall signals of 
non-collinear magnets and magnets with complex spin-textures as well 
as their  characterization through Hall experiments. 

\section*{Acknowledgments}
We are grateful to Fabian Lux and Yuriy Mokrousov for enlightening discussions 
on the different Hall contributions. We gratefully acknowledge financial support 
from the DARPA TEE program through grant MIPR (\# HR0011831554) from DOI, 
from the European Research Council (ERC) under the European Union's Horizon 
2020 research and innovation program (Grant No.\ 856538, project ``3D MAGiC'' 
and ERC-consolidator grant 681405 — DYNASORE), from Deutsche For\-schungs\-gemeinschaft 
(DFG) through SPP 2137 ``Skyrmionics'' (Project BL 444/16) and the Collaborative 
Research Centers SFB 1238 (Project C01), respectively. 

%

\end{document}